\documentclass[%
reprint,
aps,
prb,
superscriptaddress,
]{revtex4-1}

\usepackage{hyperref}
\hypersetup{colorlinks=true, linkcolor=blue, urlcolor=blue, citecolor=blue}
\usepackage{graphicx}
\usepackage{amsmath} \usepackage{amssymb}
\usepackage{accents} \usepackage{mathrsfs} \usepackage{setspace}
\usepackage{color}
\definecolor{dgreen}{RGB}{00, 120, 00} \definecolor{dblue}{RGB}{00, 00, 220}
\definecolor{lgreen}{RGB}{46, 139, 87} 
\usepackage{ulem}

\newcommand{\bs}[1]{\boldsymbol{#1}} \newcommand{\ul}[1]{\underline{#1}}

\newcommand{\ut}[1]{\undertilde{#1}} 
 \newcommand{\vphi}[0]{\varphi}

\newcommand{\ua}[0]{\uparrow} \newcommand{\da}[0]{\downarrow} 

\newcommand{\mri}[0]{\mathrm{i}} 

\begin{document}

%
%

\title{Supercurrent reversal in Zeeman-split Josephson junctions}

\author{Shu-Ichiro Suzuki}
\affiliation{MESA+ Institute for Nanotechnology, University of Twente, 
7500 AE Enschede, The Netherlands}

\author{Yasuhiro Asano}
\affiliation{Department of Applied Physics, Hokkaido University, 
Sapporo 060-8628, Japan}

\author{Alexander A. Golubov}
\affiliation{MESA+ Institute for Nanotechnology, University of Twente, 
7500 AE Enschede, The Netherlands}

\date{\today}

\begin{abstract}
We study theoretically the shape of
	the current-phase relation in a Josephson junction comprising the
	Zeeman-split superconductors (ZSs) and a normal metal (N). We show
	that at low temperatures the Josephson current in the ZS/N/ZS
	junctions exhibits an additional reversal in direction at a certain
	phase difference $\vphi_c \in (0, \pi)$. 
	Calculating the spectral Josephson current, the band-splitting due to the Zeeman interaction is shown to cause the level crossing in the spectra of the Andreev bound states and the sign reversal in the Josephson current. 
	Additionally, we propose an alternative method to electrically control
the critical phase difference $\vphi_c$ by tuning the Rashba
spin-orbit coupling, eliminating the need for manipulating
magnetizations.

\end{abstract}

\pacs{pacs}

\thispagestyle{empty}

\maketitle

\section{Introduction}

The relation between the Josephson current and the phase difference
$J(\vphi)$, the so-called current-phase relation (CPR), characterizes
Cooper-pair transport mechanisms in Josephson junctions
\cite{Golubov_RMP_04}. 
The CPR reflects well the propagation of Cooper pairs that is governed
by junction characteristics such as transmission probability of the
normal segment and the pairing symmetry of superconducting segments. 
With advances in experimental techniques, the CPR recently became a
measurable quantity using superconducting quantum interference devices
(SQUIDs).\cite{squid_PRL_2007, squid_15, squid_17, squid_17-2, ChuanChuanChuan, squids_18, squid19, squid20, squid20_2,
Fominov_PRB_22, squid3, squid4}

The typical CPR is given by $J \sim \sin \vphi$ which can be realised , for example, in the tunneling limit of the Josephson junction with the BCS-type superconductors. In the high-transparency limit, higher harmonics of the CPR are generated. In the ballistic regime, the CPR crosses over to a saw-tooth shape \cite{KO}  with a jump at $\vphi = \pm \pi$.
Recently, it was predicted that  $4 \pi$-periodic CPR might be realized at low temperatures in a Josephson junction hosting the Majorana bound states (MBSs). The Andreev bound states (ABSs),\cite{Hara,CRHu} including the MBSs,\cite{Sato_PRL_09, Luthyn_PRL_10, Oreg_PRL_10, Akz_PRB_16, Suzuki_PRB_18} stemming from the unconventional Cooper pairing\cite{Shu_PRB_14, Shu_PRR_21, Shu_PRR_22, Yoshi_PRR_22} change the transmission of the quasiparticles by the resonant tunneling.\cite{Tanaka_PRL_95, Sat_2000, Asano_PRB_04, Daghero_12, Sat_PRB_16, Agg_16, Lin_PRB_18, Shu_PRB_18, Shu_PRB_21, Ike_PRR_21} The $4 \pi$-periodic Josephson currents were recently observed in topological superconducting junctions that may demonstrate the realization of the MBSs.\cite{Molenkamp, ChuanChuanChuan, ChuanChuanChuanChuan}

At the same time, the CPR can be qualitatively modified by the
Zeeman-splitting (i.e., spin-splitting
superconductors\cite{Meservery_PRL_70, Meservery_PR_94,
Bergeret_PRL_01, Li_PRB_02, Asano_PRB_07, Giazotto_PRB_08,
TYokoyama_PRB_14, Emamipour_14, Matthias_RPP_15, Linder_NP_15,
Tatsuki_PRB_17, Maiani_arXiv_23}). The Josephson current in the
diffusive SFcFS junction has been studied using the quasiclassical
Usadel theory \cite{Golubov_JETP_02}, where S, F, and c stand for a
superconductor, ferromagnetic metal, and constriction, respectively.
It was shown that the Josephson current at low temperature
exhibits an additional reversal in direction at an intermediate phase
difference $\phi_c \in (0, \pi)$ (i.e., critical phase difference) in
addition to the standard current reversals at $\vphi=0$ and $\pi$.\cite{Golubov_JETP_02, Maiani_arXiv_23} 
In other words, the CPR at low temperature has an extra abrupt jump at 
$\vphi = \vphi_c$. 
Such an unconventional CPR, however, was not detected in experiments
yet. To observe the current reversal at $\vphi_c$, we need to
understand how to control this behavior to suggest an ideal
experimental setting. 

In this paper, the one-dimensional Josephson junctions with the
Zeeman-splitting superconductors (ZSs) are considered.  In particular,
we investigate the mechanism of the Josephson-current reversal at
$\vphi_c$ and consider an experimental setup to observe this effect. 
Using the recursive Green's function (GF) method in the lattice model,
we obtain the CPR with varying the junction parameters: magnetizations
in the ZSs, junction length, and temperature.  We have shown that the
current reversal at $\vphi_c$ appears when the magnetizations are not
antiparallel and can be the most prominent when the magnetizations are
parallel at low temperatures. Analyzing the spectral Josephson
current, we also show the origin of the
anomalous current reversal in the CPR. We
discuss the relation between the critical phase difference and the
Andreev level (i.e., the energy level of the quasiparticle bound state
at the interface). 

In addition, we demonstrate that the shape of the CPR and $\vphi_c$
can be electrically controlled by changing the Rashba spin-orbit
coupling (SOC) in the normal segment, which is easier than controlling
the magnetizations of the ZSs. The Rashba SOC effectively changes the
magnetization configuration of the junction through the spin
precession\cite{Datta_90, Manchon_15} in the normal segment.
To control
$\vphi_c$ in the absence of the Rashba SOC, one has to tune the
misalignment between two magnetizations in the ZSs. At the same time,
by tuning the SOC strength, one can qualitatively reproduce all of the
CPR types without changing the direction of the magnetizations. 

\begin{figure}[b]
	\includegraphics[width=0.48\textwidth]{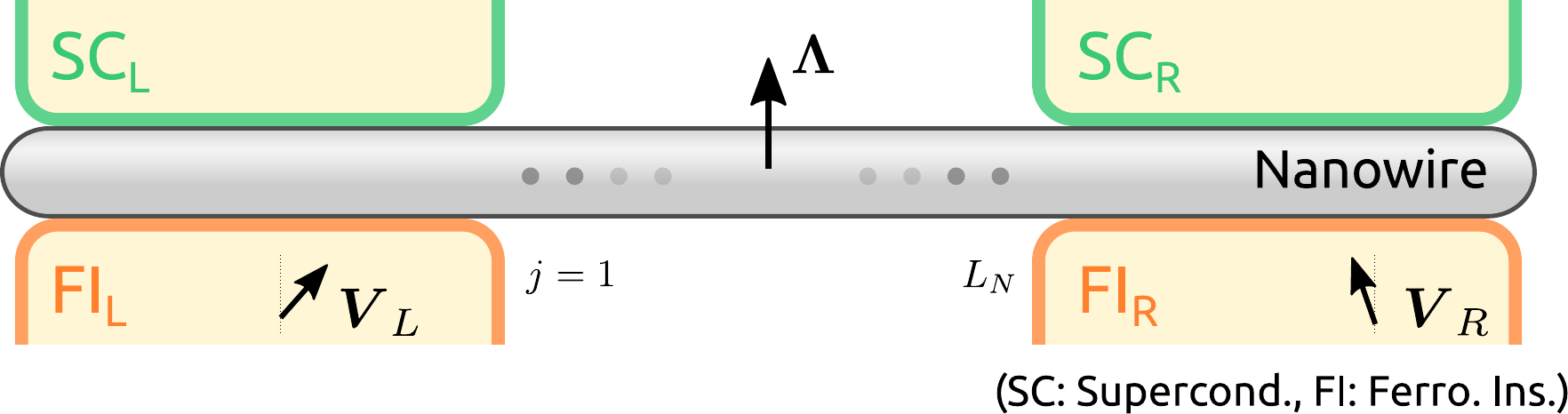}
	\caption{
		Schematic of the system. The junction consists of Zeeman-split
		superconductor (ZSs) and a nanowire with a strong
		SOC. The spin-splitting
		directions in the SCs are characterised by $V_{L(R)}$. The length
		of the nanowire is characterized by $L_N$. 
	}
	\label{fig:System}
\end{figure}

\section{Model and formulation}

We consider a one-dimensional Josephson junction with two ZSs. The two ZSs are separated by the normal segment with the length $L_N$ where the Rashba SOC is present. The  Zeeman-splitting superconducting state can be realised in the structure show in Fig.~\ref{fig:System}, where the pair potential and the Zeeman interaction are present in the wire because of the proximity from the conventional SCs and the ferromagnetic insulators.

The Hamiltonian in the normal segment is given by 
\begin{align}
  & \mathcal{H}_N
	= -t \sum_{j, \alpha}
	\left[
	    c^\dagger_{j+1,\alpha} c_{j  ,\alpha} 
	  + c^\dagger_{j  ,\alpha} c_{j+1,\alpha} 
	\right]
	\notag \\
	& \hspace{9mm}
	+i\frac{\lambda}{2} \sum_{j, \alpha, \beta}
	\left[
	    c^\dagger_{j+1,\alpha} (\hat{\sigma}_2)_{\alpha\beta}c_{j  ,\beta} 
	  - c^\dagger_{j  ,\alpha} (\hat{\sigma}_2)_{\alpha\beta}c_{j+1,\beta} 
	\right]
	\notag \\
	& \hspace{9mm}
	+ \sum_{j, \alpha}
	    c^\dagger_{j,\alpha} (2t-\mu_N) c_{j  ,\alpha}, 
\end{align}
where $t$, $\lambda$, $\mu_N$ are the hopping energy, Rashba SOC, chemical potential in the normal segment. The creation and annihilation operators are denoted by $c_{j,\alpha}$ and $c_{j,\alpha}^\dagger$ with the lattice site $j$ and the spin $\alpha$ ($\beta$). The Pauli matrices in spin and Nambu space are denoted by $\sigma_\nu$ and $\tau_\nu$ with $\nu \in \{1,2,3\}$, respectively.  The identity matrix in each space is defined as $\sigma_0$ and $\tau_0$.  In this paper, the accents $\hat{\cdot}$ and $\check{\cdot}$ mean the $2 \times 2$ and $4 \times 4$ matrices in the spin and Nambu space. 
The Hamiltonian in the superconducting lead wires are 
\begin{align}
  & \mathcal{H}_{\mathrm{i}}
	= -t \sum_{j, \alpha}
	\left[
	    c^\dagger_{j+1,\alpha} c_{j  ,\alpha} 
	  + c^\dagger_{j  ,\alpha} c_{j+1,\alpha} 
	\right]
	\notag \\
	& \hspace{14mm}
	+ \sum_{j, \alpha}
	    c^\dagger_{j,\alpha} \left[
			(2t-\mu_S) \hat{\sigma}_0 
			- \bs{V}_{\mathrm{i}}\cdot \hat{\bs{\sigma}}
			\right]_{\alpha\beta}
			c_{j  ,\beta} 
	\notag \\
	& \hspace{14mm}
	+ \sum_{j}
	\left[
	    \Delta e^{i \vphi_{\mathrm{i}}}
	    c^\dagger_{j,\ua} 
			c^\dagger_{j,\da} 
			+ \mathrm{H.c.}
			\right], 
\end{align}
where  $\mu_S$ is the chemical potential, $\Delta$ is the amplitudes of the pair potential, and the subscript $\mathrm{i} = L$ ($R$) specifies the left (right) SC. 

The electric current is obtained from the Matsubara GF $\check{\mathcal{G}}_{j,j'} (i \omega_n)$ in the normal segment, \cite{Furusaki_94, Asano_PRB_01-1, Asano_PRB_01-2}
\begin{align}
  & J= \frac{ie}{2 \hbar}T
	\sum_{\omega_n} J_{n}, \\
	& J_{n} = \mathrm{Tr} \left\{ \check{\tau}_3 \left[ 
			\check{t}_+ \check{\mathcal{G}}_{j,j+1} (i \omega_n)
	   -\check{t}_- \check{\mathcal{G}}_{j+1,j} (i \omega_n)
	\right] \right\}
\end{align}
with $\omega_n = (2n+1) \pi T$ is the Matsubara frequency, $T$ is the temperature,  and $e<0$ is the charge of the quasiparticle.  The hopping matrix is defined as 
\begin{align}
  \check{t}_\pm = 
	\left[ \begin{array}{cc}
	  \hat{t}_\pm & 0 \\
	  0 & \hat{t}^*_\pm \\
	\end{array} \right], 
	\hspace{6mm}
  \hat{t}_\pm = 
	\left[ \begin{array}{cc}
	  -t & \mp \lambda/2 \\
	  \pm \lambda/2 & t \\
	\end{array} \right]. 
\end{align}
The Josephson current can be calculated also in the real frequency representation with which we can see the relation between the Josephson current and the Andreev levels. The Josephson current is given with the spectral current $J_{E}(E)$, 
\begin{align}
  & J= \frac{e}{2 \hbar}
	\int J_{E} \tanh \left( \frac{E}{2T} \right) dE, \\
	& J_{E} =  \frac{1}{2 \pi} \mathrm{Tr} \left\{ \check{\tau}_3 \left[ 
	 \check{t}_+ \check{G}_{j,j+1} (E')
	-\check{t}_- \check{G}_{j+1,j} (E')
	\right] \right\}, 
	\label{eq:JE}
\end{align}
where $\check{G}_{j,j+1} (E')$ is the retarded GF with  $E' = E+i \delta$ with $\delta$ being the smearing factor (i.e., an infinitesimal real number). 

Throughout this paper, the Matsubara GF $\check{\mathcal{G}}_{j,j'} (i \omega_n)$ and the retarded GF $\check{G}_{j,j'}^R (E)$ are calculated by the recursive GF method \cite{Lee-Fisher}. The amplitudes of the magnetization in the ZSs are assumed to be the same ($V=V_L=V_R$ with $V_{\mathrm{i}} = |\bs{V}_{\mathrm{i}}| $), whereas the directions can be different. We assume the Zeeman interaction is smaller than $0.5 \Delta_0$ so that the pair potentials in the ZSs are finite.\cite{CC1, CC2}
The ratio between the zero-temperature pair potential and the hopping energy is set to $\Delta_0 = 0.01t$. The temperature dependence of the pair potential is calculated by the BCS relation. 
The current density is normalised to $J_0 = |e|\Delta_0/2 \hbar$, and the smearing factor is set to $\delta=0.02\Delta_0$. 

\section{Anomalous current reversal in the Josephson current}

\begin{figure}[tb]
	\includegraphics[width=0.48\textwidth]{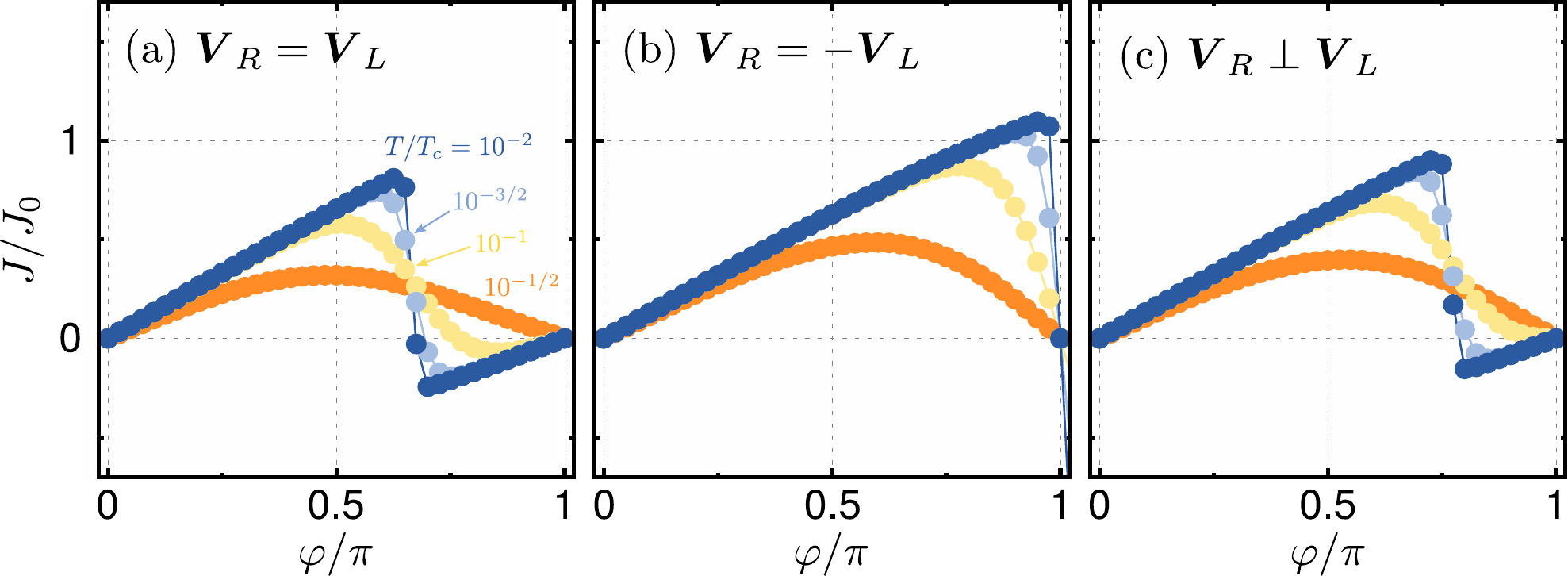}
	\caption{
	Current-phase relation in ZS/N/ZS junction.  The amplitudes of the magnetization are set to $V_L = V_R = 0.5 \Delta_0$.  The magnetization in the left is $\bs{V}_L \parallel \bs{e}_z$ and that in the right is directed to 
	(a) $\bs{e}_z$, 
	(b) $-\bs{e}_z$, and 
	(c) $\bs{e}_y$. 
	The temperature is set to $T=T_c10^{n_T}$, where $n_T$ varies from $-2$ (blue line) to -0.5 (orange line) by 0.5. The length of the junction and the chemical potential are $L_N = 80$ and $\mu_S=\mu_N=0.5t$. The legend in (a) refers to all panels. }
	\label{fig:cpr_SNS_Z}
\end{figure}
\begin{figure}[tb]
	\includegraphics[width=0.48\textwidth]{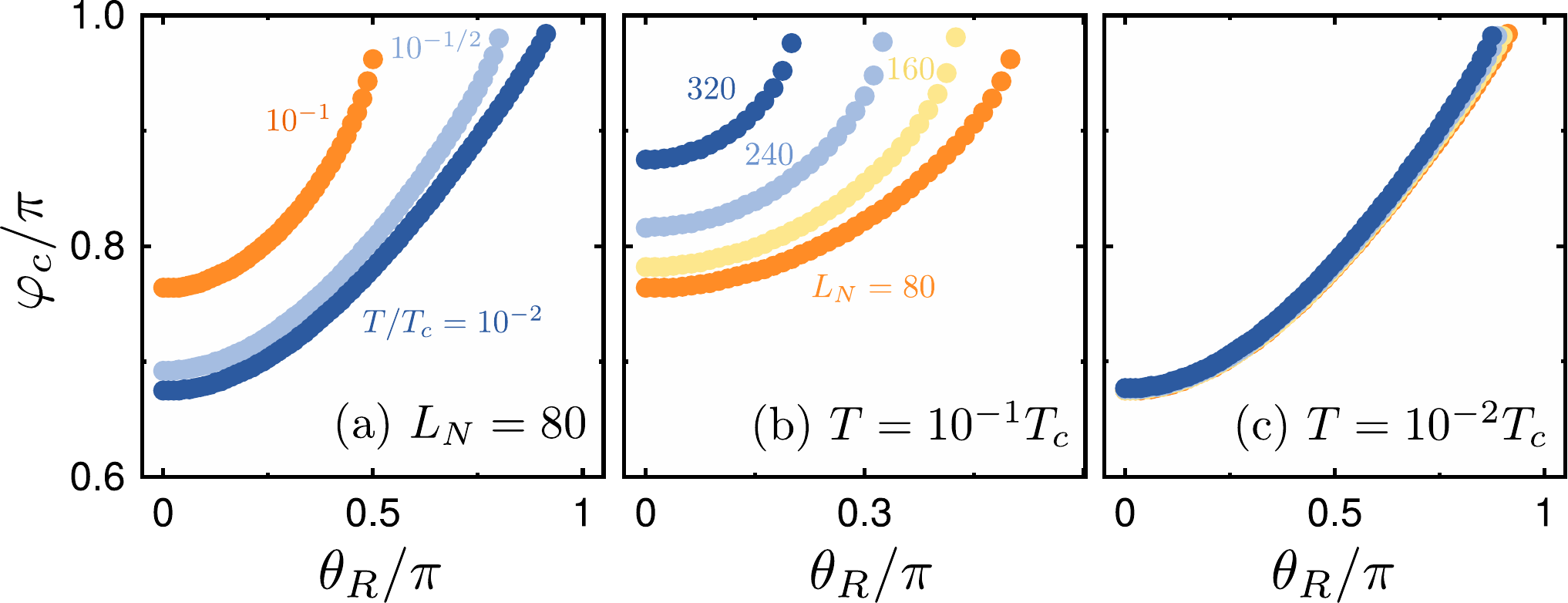}
	\caption{
		(a) Temperature dependence of critical phase in ZS/N/ZS junction.  The magnetization in the left is $\bs{V}_L \parallel \bs{e}_z$.  The magnetization in the right is $\bs{V}_R = \cos \theta_R \bs{e}_z + \sin \theta_R \bs{e}_x$. 
	  The amplitudes of the magnetization are set to $V_L = V_R = 0.5 \Delta_0$.  The temperature is set to $T=T_c10^{n_T}$, where $n_T$ varies from $-2$ to -1 by 0.5.  The length of the junction is $L_N = 80$.  
		(b,c) Length dependence of critical phase.  The temperature is set to (b) $T/T_c=10^{-1}$ and (c) $10^{-2}$.  For (c), the same legend is used as in (b).
	}
	\label{fig:phic_SNS_Z}
\end{figure}
We first show the numerical results of the CPRs \textit{without} the SOC in Fig.~\ref{fig:cpr_SNS_Z}. The amplitudes of the magnetizations are set to $V = 0.5 \Delta_0$. The magnetizations are (a) parallel, (b) antiparallel, and (c) perpendicular where the magnetization in the left SC is fixed to $\bs{V}_L \parallel \bs{e}_z$. The temperature is set to $T=T_c10^{n_T}$, where $n_T$ varies from $-2$ (blue line) to -0.5 (orange line) by 0.5. 

When the magnetizations are parallel [Fig.~\ref{fig:cpr_SNS_Z}(a)],
the Josephson currents at low
temperatures change the direction at a certain phase difference that
is not 0 nor $\pi$. We defined this phase difference as the critical
phase difference $\vphi_c \in (0, \pi)$.
Hereafter, we mainly focus on the jump in the CPR appearing
at low temperatures and the corresponding critical phase difference
that can be detected in experiments. The anomalous sign
change disappears when the magnetizations are antiparallel
[Fig.~\ref{fig:cpr_SNS_Z}(b)] where the CPR changes from $\sin \phi$
to {the saw-tooth shape \cite{KO}} as
decreasing temperature. The CPRs with parallel and antiparallel
configurations are qualitatively the same as obtained in the diffusive
limit.\cite{Golubov_JETP_02}
We can numerically obtain the CPR with non-collinear magnetizations in the recursive GF method. When the magnetizations are perpendicular, the abrupt sign change also appears as shown in Fig.~\ref{fig:cpr_SNS_Z}(c). In this case, the critical phase difference is larger than that in Fig.~\ref{fig:cpr_SNS_Z}(a). 

We show the relations between $\vphi_c$ and the misalignment of the magnetizations $\theta_R$ in Fig.~\ref{fig:phic_SNS_Z}, where $\theta_R$ is defined as $\bs{V}_R = V_L(\cos \theta_R \bs{e}_z + \sin \theta_R \bs{e}_x)$. Figure~\ref{fig:phic_SNS_Z}(a) shows that $\vphi_c$ minimizes when the magnetizations are parallel ($\theta_R=0$).  The critical phase $\vphi_c$ approaches $\pi$ with increasing $\theta_R$, meaning that the sign reversal at an intermediate phase difference disappears not only when the magnetizations are antiparallel but also the misalignment is sufficiently large. 

\begin{figure}[tb]
	\includegraphics[width=0.48\textwidth]{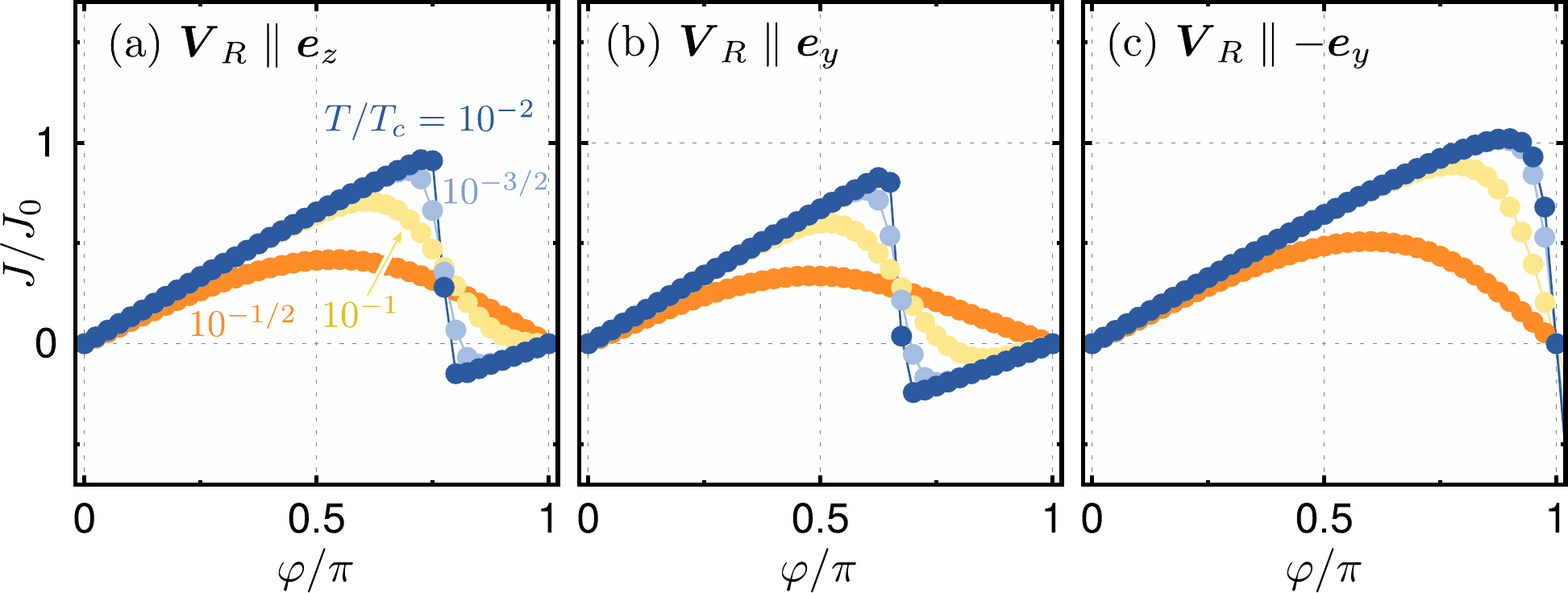}
	\caption{
	Current-phase relation in ZS/NW/ZS junction.  The SOC in the NW is set to $\lambda = 0.5t$.  The magnetization in the left is $\bs{V}_L \parallel \bs{e}_y$ and that in the right is directed to 
	(a) $\bs{e}_z$, 
	(b) $\bs{e}_y$, and 
	(c) $-\bs{e}_y$. 
	The other parameters are set to the same values as used in Fig.~\ref{fig:cpr_SNS_Z}.  The legend in (a) refers to all panels. 
	}
	\label{fig:cpr_SNS_Y}
\end{figure}

Figure~\ref{fig:phic_SNS_Z}(a) also shows that the shape of the CPR depends strongly on the temperature. We show the {$\vphi_c(\theta_R)$ curves} for different junction length $L_N$ at $T/T_c=10^{-1}$ and $10^{-2}$ in Figs.~\ref{fig:phic_SNS_Z}(b) and \ref{fig:phic_SNS_Z}(c) respectively. 
At the higher temperature [Fig.~\ref{fig:phic_SNS_Z}(b)], {$\vphi_c$} becomes larger (i.e., the {the Josephson-current reversal becomes less sharp}) because the thermal smearing diminishes the phase coherence of the quasiparticle. Even when $L_N=80$, the {anomalous current reversal} disappears when $\theta_R \sim 0.5 \pi$. In the longer junction, $\vphi_c$ becomes larger and the {non-sinusoidal behaviour}{anomalous current inversion} disappears with the smaller misalignment $\theta_R$. 

At the lower temperature [Fig.~\ref{fig:phic_SNS_Z}(c)],  the critical phase difference $\vphi_c$ is almost independent of the junction length. At this temperature regime, the coherence length (i.e., $\xi_T \sim \hbar v_F/T$) is much longer than the junction length. In other words, all of the junction in Fig.~\ref{fig:phic_SNS_Z}(c) are the short-junction regime (i.e., $\xi_T > L_N$).

\begin{figure}[tb]
	\includegraphics[width=0.48\textwidth]{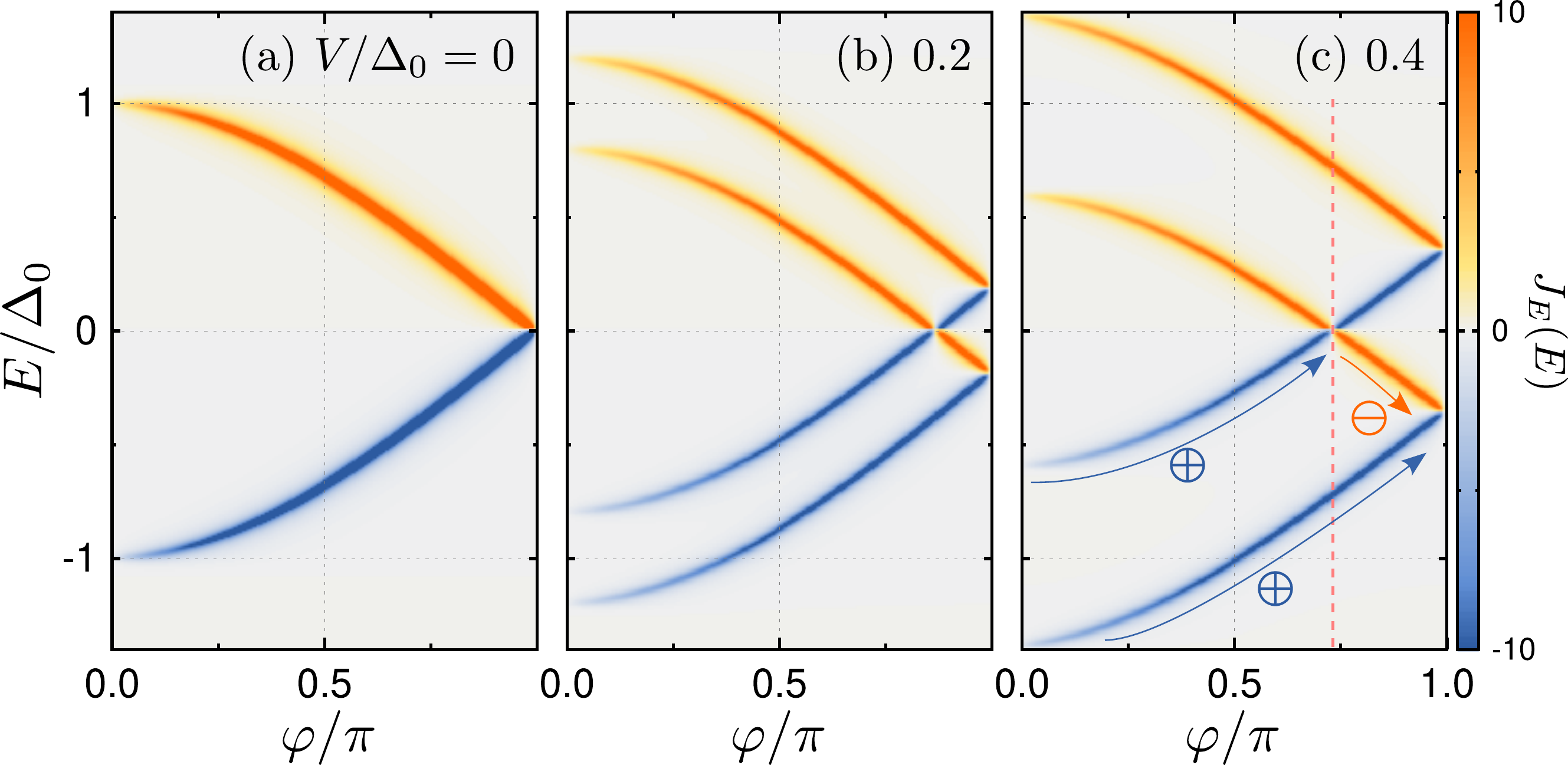}
	\caption{
	  Spectral Josephson currents in ZS/N/ZS junction \textit{without} SOC. 
		The magnetizations are set to parallel: $\bs{V}_R = \bs{V}_L = V \bs{e}_z$. The amplitude of magnetization is set to (a) $V/\Delta_0 = 0$, (b) $0.2$, and (c) $0.4$. 
		The junction length, the chemical potential, and the smearing factor  are set to $L_N=20$, $\mu=t$, and $\delta = 0.02\Delta_0$.  The signs in (c) indicate the sign of the contribution to the total current. 
	}
	\label{fig:JE_lamS0N0}
\end{figure}

The current phase relations \textit{with} the SOC are shown in Fig.~\ref{fig:cpr_SNS_Y}, where we set $\lambda = 0.5t$.  Note that we assume that one of the Zeeman interaction has the same matrix structure as that of the SOC in spin space (i.e., $\bs{V}_L \cdot \hat{\bs{\sigma}} \sim V_L \hat{\sigma}_2 \sim k \hat{\sigma}_2$).  The CPRs shown in Fig.~\ref{fig:cpr_SNS_Y} are qualitatively the same results as those in Fig.~\ref{fig:cpr_SNS_Z} (see Appendix \ref{Appen:SOC} for details). Namely, the SOC does not play an important role when one of the magnetizations is parallel to the SOC in the spin space. 

The origin of the current reversal at $\vphi_c$ can be understood by
the spin-band splitting by the Zeeman effects. The spectral Josephson
currents $J_E(E)$  [Eq.~\eqref{eq:JE}] are shown in
Fig.~\ref{fig:JE_lamS0N0}, where we fix (a) $V/\Delta_0=0$, (b) $0.2$,
and (c) $0.4$ with $\bs{V}_L \parallel \bs{V}_R$, and $L_N=20$.  Note
that, to obtain the total Josephson
current, we need to multiply the factor $\tanh(E/2T)$ to $J_E$ and
integrate on the energies.
In the absence of the magnetization, the spectral Josephson current has peaks approximately at $E=\pm \Delta_0 \cos(\vphi/2)$ that can be obtained by the Usadel theory in the ZS/constriction/ZC Josephson junction (See Appendix \ref{Appen:Usadel} for details). There are four branches in total: two spin-degenerating branches with $J_E>0$ ($J_E<0$) in the $E>0$ ($E<0$) region. 

Under the parallel magnetizations, the Zeeman interaction shifts these branches by $\pm V$ [Fig.~\ref{fig:JE_lamS0N0}(b,c)] depending on their spins. When $\vphi$ is larger than a certain $\vphi_c$ (i.e., critical phase difference defined from the CPRs), two branches crosses the zero-energy. 
The branches in $E<0$ ($E>0$)
contribute to the total current in the opposite way as indicated by
the signs in Fig.~\ref{fig:JE_lamS0N0}(c). 
This distinction is the underlying reason for the sudden change in the total current amplitude at $\vphi=\vphi_c$.

In the quasiclassical limit ($\Delta_0 \ll \mu$), we can demonstrate that the contributions from the branches cancel perfectly each among and the total current becomes zero (see Appendix \ref{Appen:Usadel} for details). However, in finite-length Josephson junctions, the total current is still finite at $\vphi_c < \vphi < \pi$ as shown in Fig.~\ref{fig:cpr_SNS_Z}(a,c). This would be because of, for example, the thermal decoherence or extra bound states trapped in the normal segment.

\section{Controlling the critical phase difference}

\begin{figure}[tb]
	\includegraphics[width=0.48\textwidth]{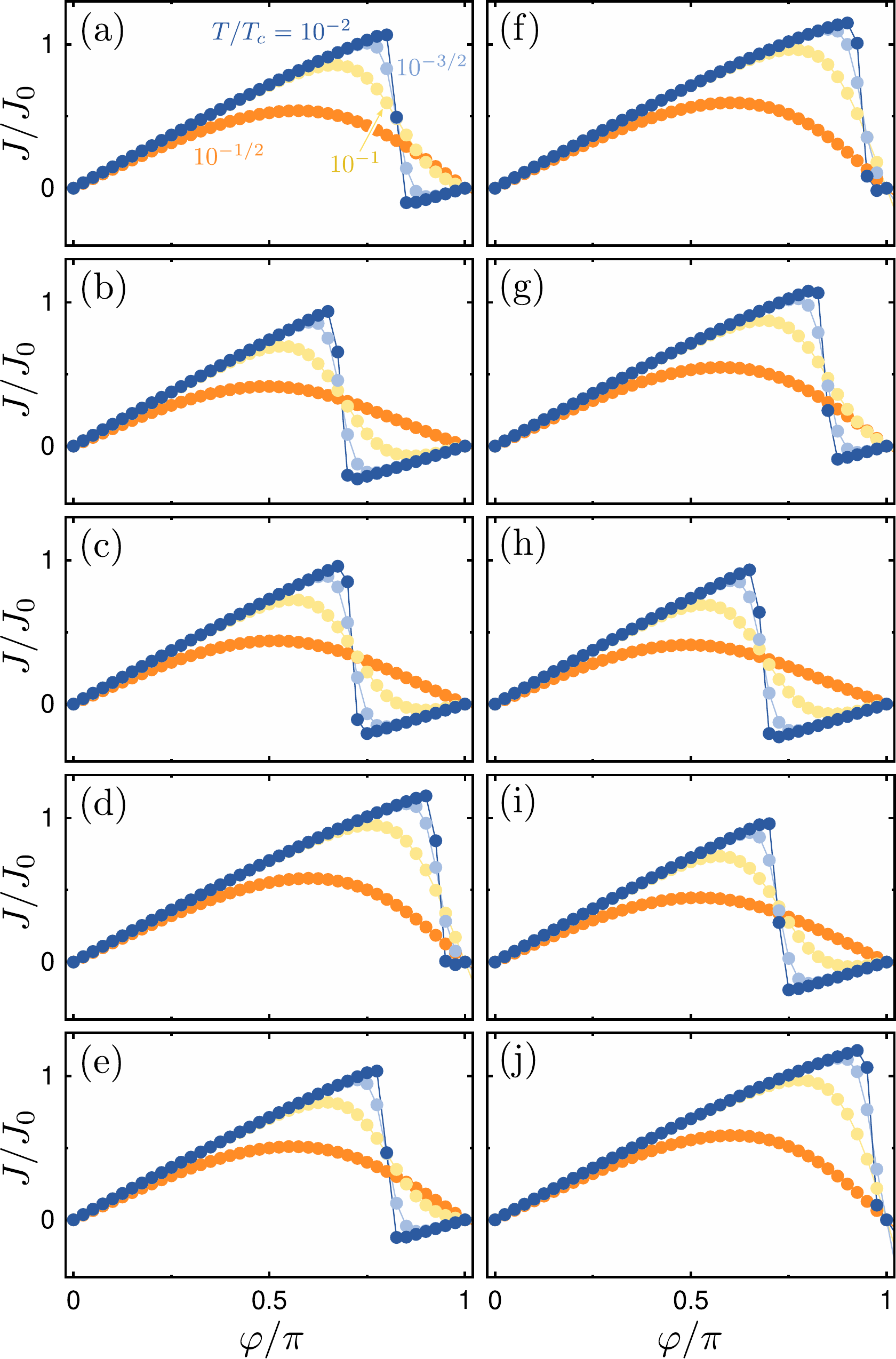}
	\caption{ (a-e) Current phase relation of ZS/NW/ZS junction with the
	Rashba SOC. The strength of the
	SOC varies from (a) $\lambda=0.5t$ to
	(e) $0.1t$ by $-0.1t$. The critical phase difference depends on
	$\lambda$.  The magnetizations are \textit{parallel} $\bs{V}_L =
	\bs{V}_R = 0.5 \Delta_0 \bs{e}_z$.  The junction length and the
	chemical potential are fixed at $L_N = 80$ and $\mu_S=\mu_N=t$. 
	(f-j) Current phase relation with \textit{perpendicular} magnetizations $\bs{V}_L = 0.5 \Delta_0 \bs{e}_z$ and $\bs{V}_R = 0.5 \Delta_0 \bs{e}_x$. These results are plotted in the same manner as in (a-e).  The legend in (a) refers to all panels.}
	\label{fig:cpr_SNWS}
\end{figure}

\begin{figure}[tb]
	\includegraphics[width=0.44\textwidth]{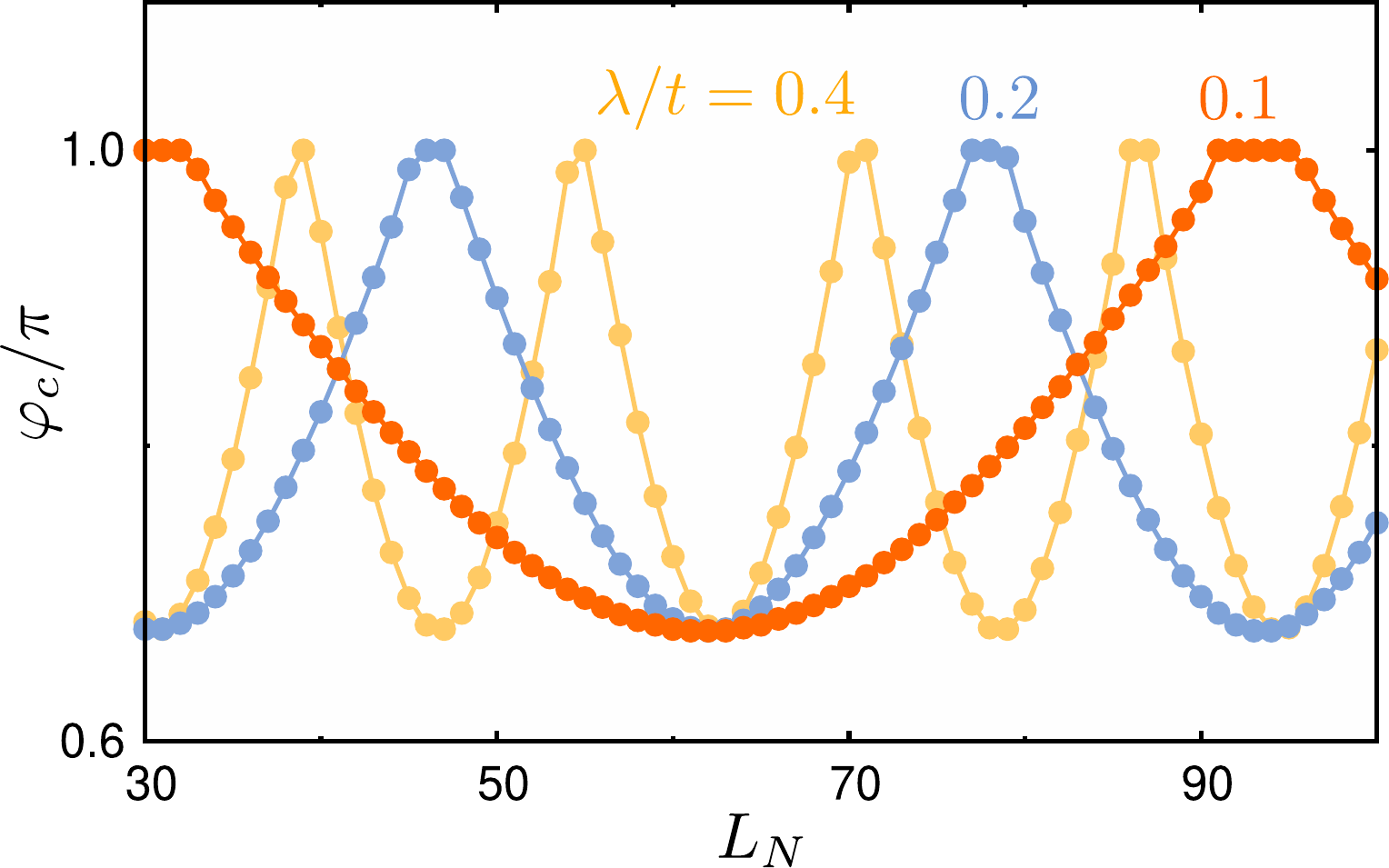}
	\caption{
		Junction-length dependences of critical phase difference in ZS/NW/ZS junction.  The magnetizations are set to  $\bs{V}_L = \bs{V}_R =0.5 \Delta_0  \bs{e}_z$.  The SOC is set to $\lambda/t=0.4$, $0.2$, and $0.1$. The temperature is set to $T=10^{-2}T_c$. 
	}
	\label{fig:phic_Ldep}
\end{figure}

\begin{figure}[tb]
	\includegraphics[width=0.48\textwidth]{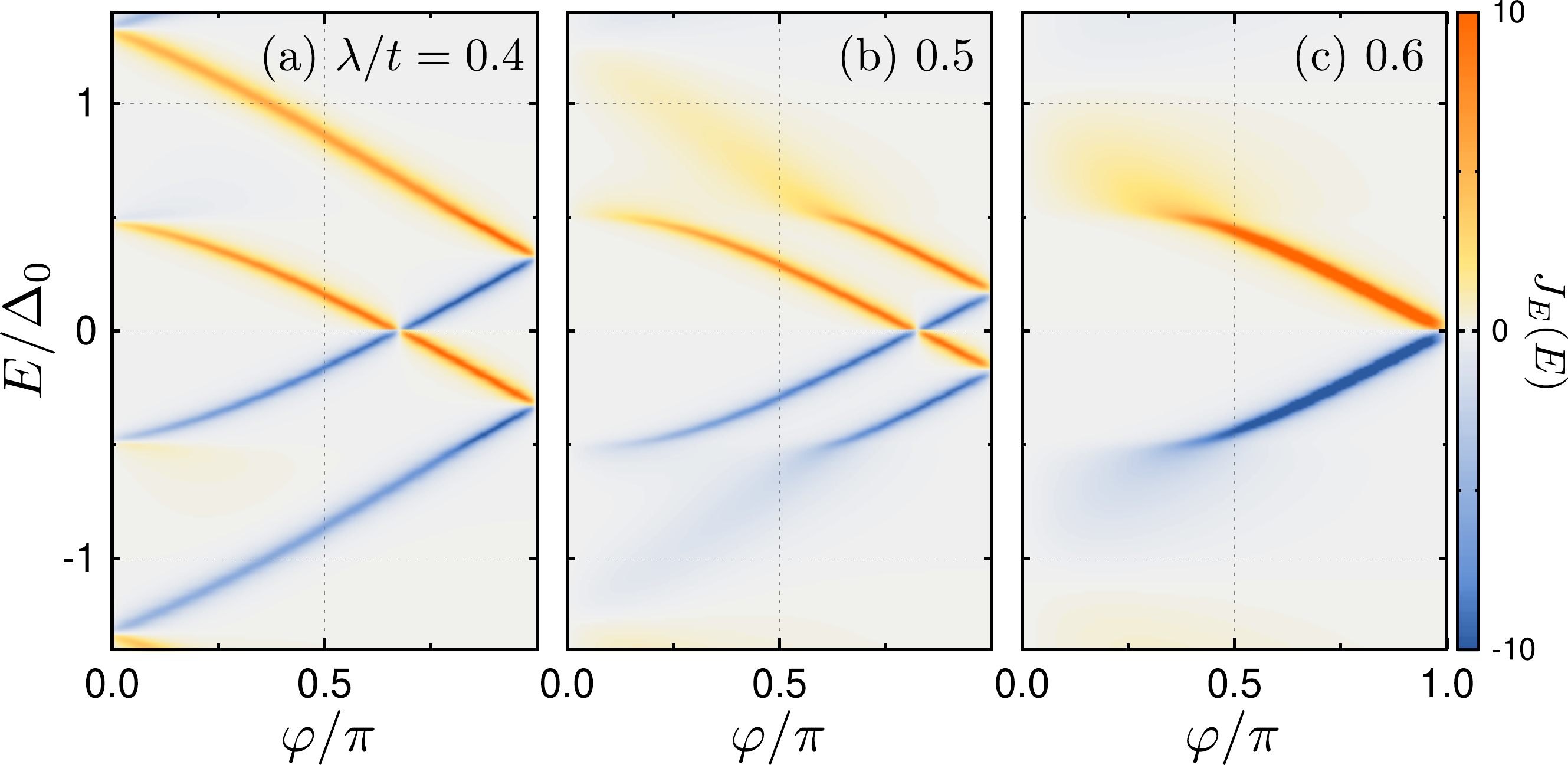}
	\caption{
	Junction-length dependence of spectral current in the presence of
	the SOC. The other parameters are set
	to the same values in Fig.~\ref{fig:cpr_SNWS}(a-e). 
	}
	\label{fig:JE_SO}
\end{figure}

When the Zeeman couplings in both superconductors have different spin
structures from the SOC, the CPR shows
a qualitatively different behavior. The CPRs with $\bs{V}_L = \bs{V}_R
= 0.5\Delta_0 \bs{e}_z$ are shown in Fig.~\ref{fig:cpr_SNWS}(a-e)
where the SOC varies from (a) $\lambda=0.5t$ to (e) $0.1t$ by $-0.1t$.
As shown in Fig.~\ref{fig:cpr_SNWS}(a-e), the critical phase
difference $\vphi_c$ changes depending on the SOC.  We also show the results with
$\bs{V}_R \parallel \bs{e}_x$ in Fig.~\ref{fig:cpr_SNWS}(f-j).
Although $\vphi_c$ is different from those in
Fig.~\ref{fig:cpr_SNWS}(a-e), $\vphi_c$ can be
controlled by $\lambda$ as in the parallel configuration. We have
confirmed this $\lambda$-dependent $\vphi_c$ never appears when
$\bs{V}_R \parallel \bs{e}_y$ (not shown) where one of the
Zeeman couplings have the same matrix structure as the SOC. 

To clarify the relation between $\vphi_c$ and $\lambda$, we show the
junction-length dependences of $\vphi_c$ for several strengths of
$\lambda$ in Fig.~\ref{fig:phic_Ldep}, where we fix $T=10^{-2}T_c$
(i.e., short-junction limit).  Figure~\ref{fig:phic_Ldep} shows
$\vphi_c$ oscillates in the junction length $L_N$. Even if the
magnetization is parallel, sign
reversal in the Josephson current can vanish as happened in the
antiparallel configuration [see Fig.~\ref{fig:cpr_SNS_Z}(b)].
Looking at the results in
Fig.~\ref{fig:phic_Ldep}, we see that the period of oscillation is
approximately proportional to
$\lambda^{-1}$. 

The oscillating behaviour is related to the spin precession by the SOC as discussed in Ref.~\onlinecite{Tatsuki_PRB_17}.  The SOC in the nanowire acts on the quasiparticle spin as an effective Zeeman field and causes the spin precession \cite{Datta_90, Manchon_15, Tatsuki_PRB_17}. Corresponding to the SOC, the quasiparticles obtain an additional phase depending on their spin when they travel across the junction. This additional phase can reproduce the situation with the antiparallel magnetizations in the non-SOC junction, where the sudden jump never appears in the CPR [Compare Figs.~\ref{fig:cpr_SNWS}(d) and \ref{fig:cpr_SNS_Z}(b)].

The SOC affects the spectral current as well.  The spectral currents
are shown in Fig.~\ref{fig:JE_SO}, where $L_N=80$ and the SOC varies
from $\lambda/t=0.4$ to $0.6$ by $0.1$. When $\lambda/t=0.4$, the band
splitting is almost maximized which results in the smallest $\vphi_c$
among the three panels in Fig.~\ref{fig:JE_SO} [see, for example,
Fig.~\ref{fig:phic_Ldep}]\footnote{Note that the width of
the band splitting shows a periodic behavior. The results in
Figs.~\ref{fig:phic_Ldep}(a) and \ref{fig:phic_Ldep}(c) correspond to
one of the minima and maxima}.  With increasing the
SOC, the band splitting diminishes and
becomes almost zero when $\lambda/t=0.6$. As a consequence, the
anomalous current reversal at an
intermediate phase difference completely disappeared. 

The magnitude of the band-splitting in the spectral current is consistent with those obtained in the continuum model.\cite{Tatsuki_PRB_17} In the continuous limit, the band splitting is estimated as $\delta E = V \cos( \lambda L_N / \hbar v_F)$, where we have made $\hbar$ explicit to avoid misunderstandings.  From this relation, we can estimate the period of the oscillation in Fig.~\ref{fig:phic_Ldep} as $L_0 \sim 5.44t/\lambda$ which is almost consistent with our numerical simulation.  Note that the continuum model is used in Ref.~[\onlinecite{Tatsuki_PRB_17}], whereas we use the tight binding model. Therefore, the estimation and our numerical result in Fig.~\ref{fig:phic_Ldep} are slightly different. In the estimation, the length $L_0$ is measured in the unit of the lattice constant and we have used $\mu=t$. 

\section{Conclusion}
We have shown that the Josephson junction with the Zeeman-splitting
superconductors (ZSs) can have a anomalous current
reversal at low temperature in the current-phase relation
(CPR) at the critical phase difference
$\vphi_c \in (0, \pi)$.  
At low temperatures, in particular, the Josephson current changes
suddenly its direction at
$\vphi_c$.  Changing the misalignment between the
magnetization in the ZSs, we have shown that
$\vphi_c$ depends on the misalignment angle between
the two magnetizations. 
Notably, the most prominent current jump is observed in the parallel
configuration, while the reversal at an intermediate phase difference
is absent in the antiparallel configuration.

Analysing the spectral Josephson current, we have shown that the
anomalous current reversal stems from the
spin-band splitting by the Zeeman interaction.  When the
magnetizations are not antiparallel, the Zeeman interaction results in
the spin-band splitting in the Andreev levels. The Josephson current
changes direction when one of the Andreev levels crosses zero energy.
However, with the antiparallel configuration, the current reversal
disappears because opposite magnetizations does not split the
spin-band of the Andreev levels. 

In addition, we have proposed the method to observe the
current reversal at $\vphi_c$.  We have
demonstrated that the $\vphi_c$ can
be electrically controlled by tuning the Rashba spin-orbit coupling
even without changing the magnetizations.  The spin-orbit coupling
causes the spin precession which determines the
width of the spin-band splitting and
the critical phase difference $\vphi_c$.  Namely, one can control
$\vphi_c$ by, for example, the gate voltage that tunes the Rashba
spin-orbit coupling.

\begin{acknowledgments}
The authors are grateful to Ya.~V.~Fominov and C.~Li for useful discussion.
S.-I.~S. acknowledges Overseas Research Fellowships by JSPS and the hospitality at the University of Twente.
This work was supported by JSPS KAKENHI (No. JP20H01857), 
JSPS Core-to-Core Program (No. JPJSCCA20170002), 
and JSPS and Russian Foundation for Basic Research under 
Japan-Russia Research Cooperative Program
(Nos. JPJSBP120194816 and 19-52-50026). 
\end{acknowledgments}

\appendix

\section{Critical phase difference with spin-orbit coupling} \label{Appen:SOC}

\begin{figure}[tb]
	\includegraphics[width=0.48\textwidth]{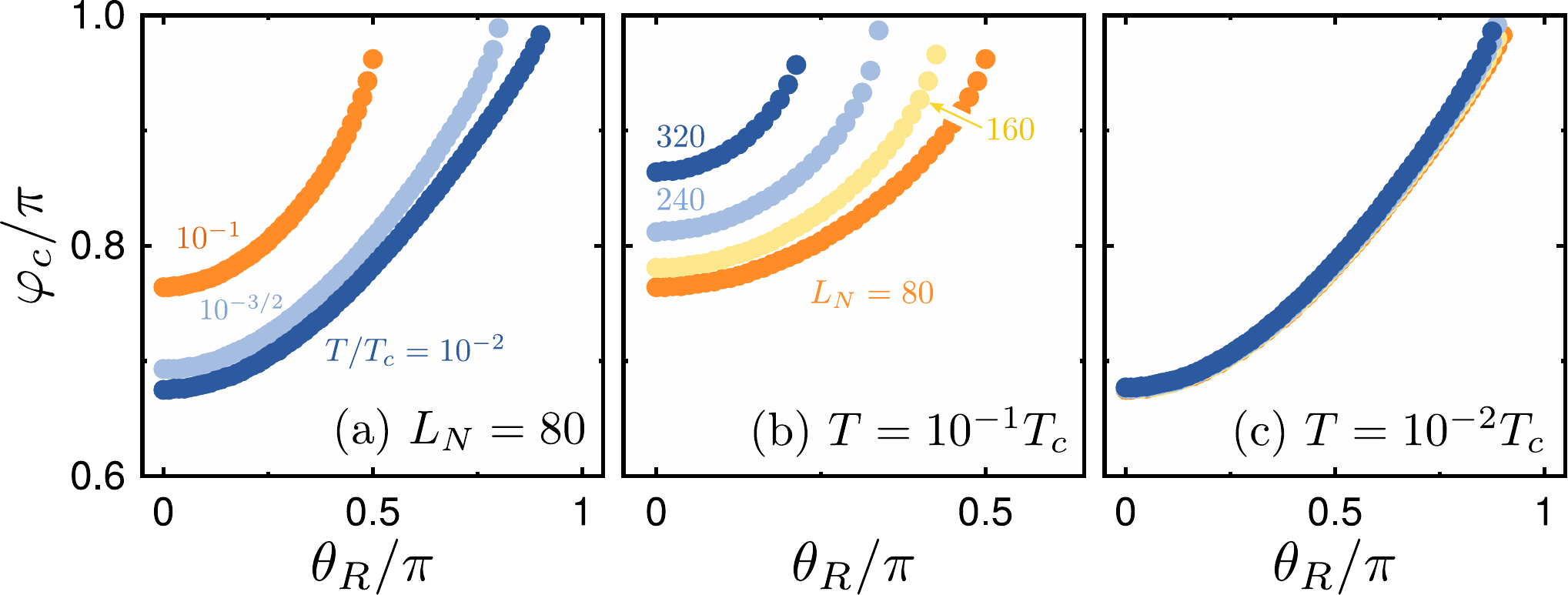}
	\caption{
		(a) Temperature dependence of critical phase in ZS/NW/ZS junction. 
		(b,c) Length dependence of critical phase. 
    The magnetization in the left is $\bs{V}_L \parallel \bs{e}_y$.
		The magnetization in the right is 
		$\bs{V}_R = \cos \theta_R \bs{e}_y + \sin \theta_R \bs{e}_z$. 
		The results are shown in the same manner as in Fig.~\ref{fig:phic_SNS_Z}. 
	}
	\label{fig:phic_SNS_Y}
\end{figure}

In this section, we discuss $\vphi_c$ in the presence of the Rashba
SOC. Note that one of the Zeeman interactions has the same matrix
structure as that of the Rashba SOC (i.e., $\bs{V}_\mathrm{i} \cdot
\hat{\bs{\sigma}} \sim V_\mathrm{i} \hat{\sigma}_2 \sim k
\hat{\sigma}_2$). The results are shown in Fig.~\ref{fig:phic_SNS_Y}
in the same manner as in Fig.~\ref{fig:phic_SNS_Z}, where the results
without the SOC are shown.  Figure~\ref{fig:phic_SNS_Y} shows that,
when the one of the Zeeman interactions is proportional to
$\hat{\sigma}_y$, the SOC with $k_x \hat{\sigma}_y$ does not
qualitatively change $\vphi_c$. 
 
\section{Gor'kov theory and anomalous current reversal in the Josephson current}
 \label{Appen:Usadel}

\subsection{Usadel theory}

In a diffusive superconducting system in equilibrium, 
the Usadel quasiclassical Green's
functions satisfy the Usadel equation: 
\begin{align}
  \hbar D \bs{\nabla} \cdot
  \big( \check{g} \bs{\nabla} \check{g} \big)
	+ i \big[i \omega_n \check{\tau}_3 + \check{H},\check{g} \big]_- = 0, 
	\label{eq:Usadel01}
\end{align}
where $D$ is the diffusion constant and $\check{g} = \check{g}(\bs{r}, i \omega_n)$ is the Matsubara
Green's function in the Nambu space (i.e., particle-hole $\otimes$
spin space) defined as, 
\begin{align}
	& \check{g}
	= \left( \begin{array}{cc}
		\hat{g} & 
		\hat{f}_{ \omega} \\[2mm]
	  \hat{f}_{-\omega}^\dagger & 
		-\ut{\hat{g}} \\
	\end{array} \right). 
	\label{eq:Usa_GF}
\end{align}
In Eq.~\eqref{eq:Usa_GF}, we have used the symmetry of the Green's
function: $-\ut{\hat{f}}_\omega =\hat{f}_{-\omega}^\dagger$ with 
the undertilde functions defined as 
$
\ut{\hat{X}}(\bs{r}, i \omega_n) = 
   {\hat{X}}(\bs{r}, i \omega_n)^*$ with $X$ being an arbitral function. 

In this note, we assume the Zeeman and the $s$\,-wave spin-singlet
pair potentials. The $\check{H}$-matrix in this
case is given by 
\begin{align}
	\check{H}
	= \left[ \begin{array}{cc}
	          \hat{\xi} &     \hat{\eta} \\[2mm]
	  \ut{\hat{\eta}} & \ut{\hat{\xi}} \\
	\end{array} \right]
	= \left[ \begin{array}{cc}
			-V \hat{\sigma_3} &
			i\Delta_0 (i\hat{\sigma}_2) \\[2mm]
			i \Delta^*_0 (i \hat{\sigma}_2)^\dagger &
			-V \hat{\sigma_3} \\
	\end{array} \right], 
\end{align}
where we have assumed the Zeeman potential $V(x)$ is in the
$\sigma_3$-direction. 
In this case, it is convenient to apply the unitary transform, 
\begin{align}
	& \check{U}^{-1} (i\omega_n \check{\tau}_3 + \check{H} ) \check{U}
	\\
	&= \left[ \begin{array}{cc}
			i \omega_n -V \hat{\sigma}_3 &
			i \Delta_0   \hat{\sigma}_0 \\[2mm]
			i \Delta^*_0 \hat{\sigma}_0 &
			-i\omega_n+V \hat{\sigma}_3 \\
	\end{array} \right], 
\end{align}
with $\check{U} = \mathrm{diag}[\hat{\sigma}_0, -i \hat{\sigma}_2]$. Accordingly, 
the Green's function can be parametrised as, 
\begin{align}
	& \check{U}^{-1} \check{g} \check{U}
	= \left( \begin{array}{cccc}
			g_+ & 0 & f_{ \omega, +} & 0 \\[2mm]
			0 & g_- & 0 & f_{ \omega, -} \\[2mm]
	  	f^*_{-\omega, +} & 0 & - \ut{g}_- & 0 \\[2mm]
			0 & f^*_{-\omega, -} & 0 & -\ut{g}_+ \\
	\end{array} \right). 
\end{align}
where we have used $\hat{g} = \mathrm{diag}[ g_+, g_-]$ and 
$\hat{f}_\omega = \mathrm{diag}[ f  _{ \omega, +}, f  _{ \omega, -}] (i
\hat{\sigma}_2)$. 
Therefore, in the following, we treat this spin-reduced
Green's function in each spin subspace, 
\begin{align}
  \tilde{g}_\sigma 
	= \left( \begin{array}{cc}
			g_\sigma & 
			f  _{ \omega, \sigma} \\[2mm]
	  	f^*_{-\omega, \sigma} & 
			- \ut{g}_{\bar{\sigma}} \\
	\end{array} \right)
	= \left( \begin{array}{cc}
			g_\sigma & 
			f  _{ \omega, \sigma} \\[2mm]
			\ul{f}_{\omega, \sigma} & 
			- \ut{g}_{\bar{\sigma}} \\
		\end{array} \right), 
\end{align}
where $\sigma = \pm$ (with $\bar{\sigma} = - \sigma$) specifies the spin subspace. 
We have introduced the underline accent as $\ul{f}_{\omega} =
f^*_{-\omega}$. 
The normalization condition becomes
\begin{align}
			g_\sigma^2 + 
			f \ul{f} 
			= 1. 
\end{align}
In the homogeneous limit, the Green's functions satisfy
\begin{align}
  & \tilde{g}_\sigma
	= \frac{1}{\Omega_\sigma}
	\left[ \begin{array}{cc}
			\omega_\sigma &
			\Delta_0    \\[2mm]
			\Delta^*_0  &
			-\omega_\sigma \\
	\end{array} \right], 
\end{align}
with  $\Omega_\sigma = \sqrt{\omega_\sigma^2 + |\Delta_0|^2}$ and 
$\omega_\sigma = \omega_n+ i \sigma V $. 


\subsection{Josephson current}

\begin{figure}[tb]
\centering
\includegraphics[width=0.48\textwidth]{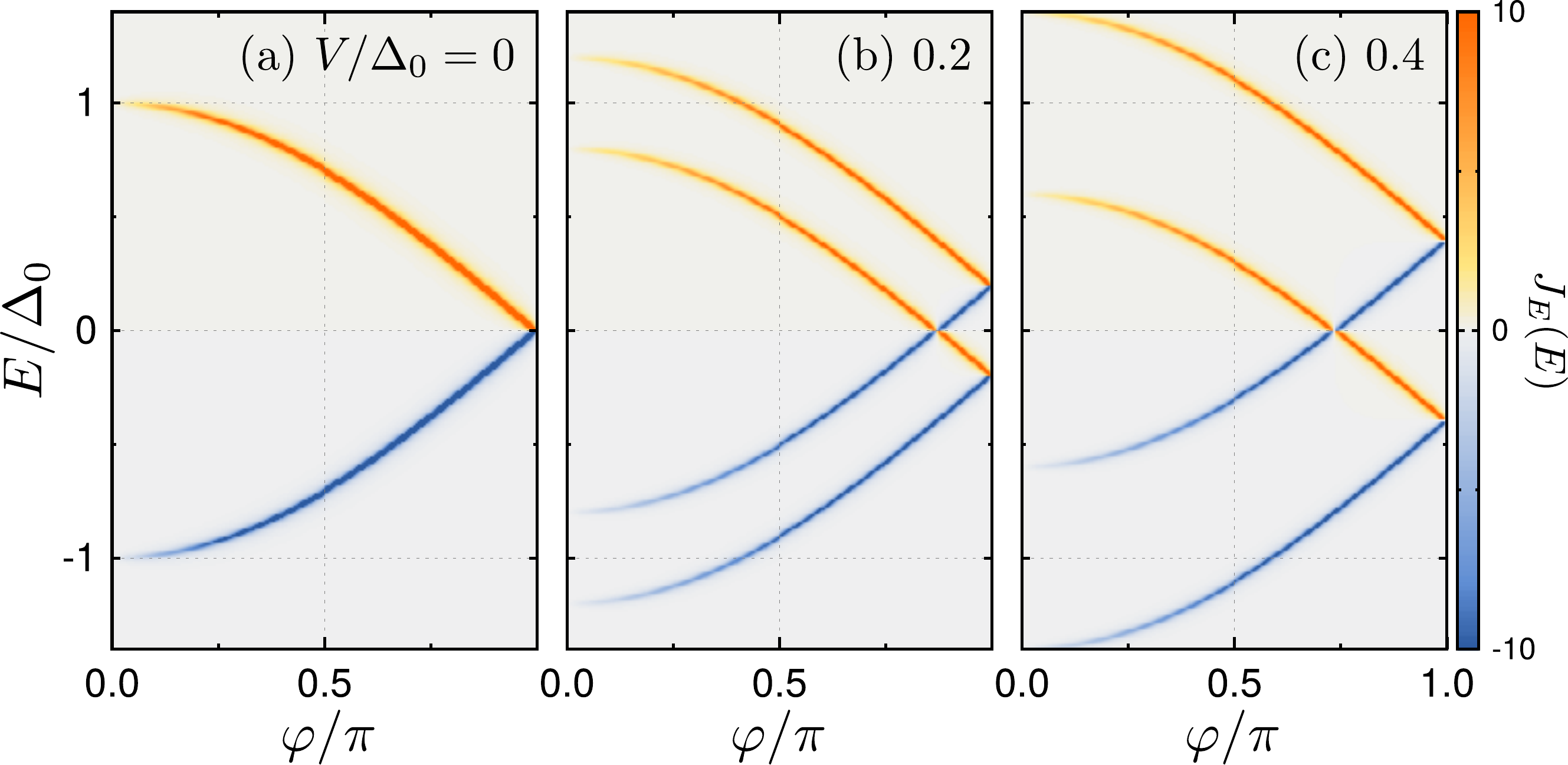}
\caption{Spectral Josephson current in ScS junction obtained by the
Usadel theory. }
\label{fig:UsaPhi}
\end{figure}

In this section, we consider the
superconductor/constriction/superconductor junction as discussed, for
example, in Ref.~\onlinecite{Golubov_JETP_02}. 
The Josephson current in an ScS junction can be written as, 
\begin{align}
  & J = 
	\frac{\pi T}{ i |e| R_N}
	\sum_{\omega_n, \sigma} J_\sigma( i\omega_n), 
	\\
	& J_\sigma
	=
	\frac{(\ul{f}_{L,\sigma} f_{R,\sigma} - f_{L,\sigma} \ul{f}_{R,\sigma})/2}
	{2-D [1-g_{L,\sigma} g_{R,\sigma} - (\ul{f}_{L,\sigma} f_{R,\sigma} + f_{L,\sigma} \ul{f}_{R,\sigma})/2]}. 
  \label{}
\end{align}
where the subscript $\mathrm{i} = L$ ($R$) specifies the left (right)
superconductor. \cite{Zaitsev}

Assuming the Zeeman-splitting superconductors, the Green's functions
are given by 
\begin{align}
  g_{\mri,\sigma} =           \omega_{\mri,\sigma} / \Omega_{\mri,\sigma}, \hspace{9mm}
  f_{\mri,\sigma} = \Delta e^{i \vphi_\mri} / \Omega_{\mri,\sigma}, 
\end{align}
where $\omega_{\mri,\sigma} = \omega_n + i \sigma V_{\mri}$, 
$\Omega_{\mri,\sigma} = 
\sqrt{\omega_{\mri,\sigma}^2 - \Delta_0^2}$. 
The current across the junction can be obtained as 
\begin{align}
  & I = 
	\frac{\pi T}{i |e| R_N}
	\sum_{\omega_n, \sigma} J_\sigma (i \omega_n), 
	\\
  & J_\sigma (i \omega_n) = 
	\frac{i \Delta_0^2 \sin \vphi}
	{2 \Omega_L \Omega_R 
	- D [\Omega_L \Omega_R - \omega_L \omega_R - \Delta_0^2 \cos \vphi]}. 
  \label{}
\end{align}
where we have omitted $\sigma$. 
When the magnetizations are parallel ($V_L = V_R$), 
the current density is reduced to 
\begin{align}
  I &= 
	\frac{2 \pi T}{|e| R_N}
	\sum_{\omega_n>0}
	\frac{ A \Delta_0^2 \sin \vphi}
	{A^2 + 4 \omega_n^2 V^2} ,
	\\
	A&=(\omega_n^2 - V^2) + \Delta_0^2 [1-D \sin^2 (\vphi/2)]. 
\end{align}
In this expression, the factor $A$ has a sign change at
\begin{align}
	 \sin^2 (\vphi_c/2) = 
	 \frac{\Delta_0^2-V^2+\omega_n^2}
	 {D \Delta_0^2}
\label{}
\end{align}
meaning that the Josephson current suddenly changes the direction at a
certain phase difference $\vphi_c$ (i.e., {anomalous 
current reversal in the Josephson current}). 
In the antiparallel junction ($V_L = -V_R$), the CPR is a standard one
(i.e., no jump appears in the CPR) because the extra phases from the
Zeeman effects cancel each other; $\omega_R = \omega_L^*$ and
$\Omega_R = \Omega_L^*$. 
  
Using the analytic continuation, we can obtain the expression of the
current in the real-frequency representation. When the magnetizations
are parallel, the current in terms of
the spectral current is given by
\begin{align}
  & J = 
	\frac{\pi}{2|e| R_N}
	\int J_E \tanh \left( \frac{E}{2T} \right) dE, 
	\\
  & J_E = -\frac{1}{2 \pi} (J^R-J^A), 
	\\
	& J^R = \sum_\sigma \frac{i \Delta_0^2 \sin \vphi}
	{\Delta_0^2 [1-D \sin^2 (\vphi/2)] 
	- {(\bar{E} - \sigma V)^2}}. 
  \label{}
\end{align}
where $\bar{E} = E+i\delta$. Using the relation $J^A = -(J^R)^*$, we
have 
\begin{align}
	& J_E = \sum_\sigma J_{E, \sigma},
	\\
	& 
	J_{E, \sigma} = 
	{
	\frac{1}{2 \pi}
	\frac{ \delta_1}
	{ \Delta_1^2 + \delta_1^2}
	\mathrm{sgn}[E-\sigma V] \Delta_0^2 \sin \vphi , 
	}
  \label{eq:LDirac}
	\\
  & \Delta_1 = \Delta_0^2 [1-D \sin^2 (\vphi/2)]-(E-\sigma V)^2, 
\end{align}
where $\delta_1 = {2|E-\sigma V|\delta}$. 
Equation \eqref{eq:LDirac} become 
the Lorentzian-type Dirac function at $\delta_1 \to 0$ that has peaks at 
\begin{align}
  E = \pm \sqrt{1-D\sin^2 ({\vphi}/2)} + \sigma V, 
  \label{}
\end{align}
which can reproduce the results in the high-transparency limit in the
absence and presence of the Zeeman field.\cite{Kulik, Tatsuki_PRB_17}
The peak positions are shifted by the Zeeman interaction in the
superconductors. The spectral currents $J_E$ are shown in
Fig.~\ref{fig:UsaPhi}, where 
(a) $V/\Delta_0=0$, 
(b) $0.2$, and 
(c) $0.4$. The results are qualitatively the same as those in
Fig.~\ref{fig:JE_lamS0N0}.


\end{document}